# A Declarative Recommender System for Cloud Infrastructure Services Selection

Miranda Zhang[1], Rajiv Ranjan[1], Surya Nepal[1], Michael Menzel[2], Armin Haller[1]

[1] Information Engineering Laboratory, CSIRO ICT Centre
{miranda.zhang, rajiv.ranjan, surya.nepal, armin.haller}@csiro.au
[2] Karlsruhe Institute of Technology, Karlsruhe, Germany
menzel@fzi.de

**Abstract.** The cloud infrastructure services landscape advances steadily leaving users in the agony of choice. Therefore, we present CloudRecommender, a new declarative approach for selecting Cloud-based infrastructure services. CloudRecommender automates the mapping of users' specified application requirements to cloud service configurations. We formally capture cloud service configurations in ontology and provide its implementation in a structured data model which can be manipulated through both regular expressions and SQL. By exploiting the power of a visual programming language (widgets), CloudRecommender further enables simplified and intuitive cloud service selection. We describe the design and a prototype implementation of CloudRecommender, and demonstrate its effectiveness and scalability through a service configuration selection experiment on most of today's prominent cloud providers including Amazon, Azure, and GoGrid.

## 1 Introduction

Cloud computing [1,2,3] assembles large networks of virtualized services: infrastructure services (e.g., compute, storage, network, etc.) and software services (e.g., databases, message queuing systems, monitoring systems, load-balancers, etc.). It embraces an elastic paradigm in which applications establish on-demand interactions with services to satisfy required Quality of Service (QoS) including cost, response time and throughput. However, selecting and composing the right services meeting application requirements is a challenging problem.

Consider an example of a medium scale enterprise that would like to move its enterprise applications to cloud. There are multiple providers in the current cloud landscape that offer infrastructure services in multiple heterogeneous configurations. Examples include, Amazon [10], Microsoft Azure [12], GoGrid [13], Rackspace, BitCloud, and Ninefold, among many others. With multiple and heterogeneous options for infrastructure services, enterprises are facing a complex task when trying to select and compose a single service type or a combination of service types. *Here we are concerned with simplifying the selection and comparison of a set of infrastructure service offerings for hosting the enterprise applications and corresponding dataset, while meeting multiple criteria, such as specific configuration and cost, emanating from the enterprise's QoS needs.* This is a challenging problem for the enterprise and needs to be addressed.

Existing approaches in helping a user to compare and select infrastructure services in cloud computing involve manually reading the provider documentation for finding out which services are most suitable for hosting an application. This problem is further aggravated by the use of non-standardized naming terminologies used by cloud providers. For example, Amazon refers to compute services as EC2 Compute Unit, while GoGrid refers to the same as Cloud Servers. Furthermore, cloud providers typically publish their service description, pricing policies and Service-Level-Agreement (SLA) rules on their websites in various formats. The relevant information may be updated without prior notice to the users. Hence, it is not an easy task to manually obtain service configurations from cloud providers' websites and documentations (which are the only sources of information).

In order to address the aforementioned problems, we present a semi-automated, extensible, and simplified approach and system for cloud service selection, called CloudRecommender. We indentify and formalize the domain knowledge of multiple configurations of infrastructure services. The core idea in CloudRecommender is to formally capture the domain knowledge of services using a declarative logic-based language, and then implement it in a recommender service on top of a relational data model. Execution procedures in CloudRecommender are transactional and apply well-defined SQL semantics for querying, inserting, and deleting infrastructure services' configurations. The CloudRecommender system proposed in this paper leverages the Web-based widget programming technique that transforms drag and drop operations to low-level SQL transactions. The contributions of this paper can be summarized as follows:

- A unified and formalized domain model capable of fully describing infrastructure services in cloud computing. The model is based and has been successfully validated against the most commonly available infrastructure services including Amazon, Microsoft Azure, GoGrid, etc.
- An implementation of a design support system (CloudRecommender) for the selection of infrastructure cloud service configurations using transactional SQL semantics, procedures and views. The benefits to users of CloudRecommender include, for example, the ability to estimate costs, compute cost savings across multiple providers with possible tradeoffs and aid in the selection of cloud services.
- A user-friendly service interface based on widgets that maps user requirements based on form inputs to available infrastructure services, express configuration selection criteria and view the results.

The remainder of the paper is organized as follows. A discussion on our formal domain model for cloud infrastructure services and our cloud selection approach using CloudRecommender is presented in Section 2. Details on the experimental evaluation of the proposed approach and system are given in Section 3. A review of related work is provided in Section 4 before we conclude in Section 5.

## 2 A System for Cloud Service Selection

We propose an approach and system for cloud service configuration selection, CloudRecommender. The system includes a repository of available infrastructure services from different providers including compute, storage and network services, as shown in figure 1(a). Users can communicate with the system via a Web-based widget interface. The CloudRecommender system architecture consists of three layers: the configuration management layer, the application logic layer and the User interface (widget) layer. Details of each layer will be explained in the following sub-sections.

Figure 1(b) shows the deployment structure of the CloudRecommender system. For persistence we have chosen MySQL for its agility and popularity, but any other relational database can be plugged in. Furthermore, many APIs provided by cloud providers (such as Amazon) and open source cloud management frameworks (e.g. jclouds) are written in Java. Thus, Java is chosen as the preferred language to implement the application logic layer to ease the integration with external libraries. The widget layer is implemented using a number of JavaScript frameworks including jQuery, ExtJS and YUI. CloudRecommender also exposes RESTful (REpresentational State Transfer) APIs (application programming interface) that help external applications to programmatically compose infrastructure cloud services based on the CloudRecommender selection process.

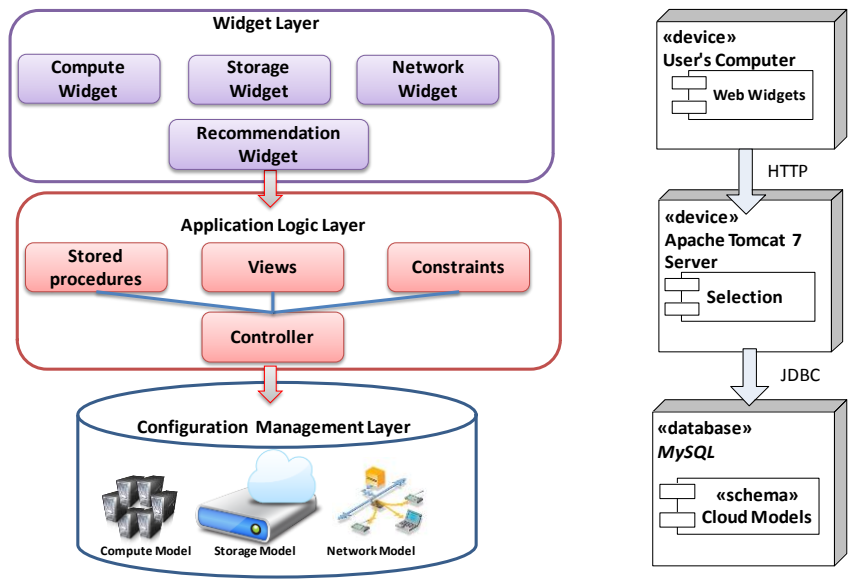

**Figure 1. (a) System architecture.            (b) Deployment structure.**

### 2.1 Configuration Management Layer

The configuration layer maintains the basic cloud domain model related to compute, storage, and network services. We defined a Cloud Computing Ontology to facilitate the discovery of services based on their functionality and QoS parameters. The ontology is defined in the Web Ontology Language (OWL) [19] and can be found at: w3c.org.au/cocoon.owl. All common metadata fields in the ontology like Organisation, Author, First Name etc. are referenced through standard Web Ontologies (i.e. FOAF and Dublin Core). To describe specific aspects of cloud computing, established domain classifications have been used as a guiding reference [16, 18]. The resulting ontology consists of two parts, the Cloud Service Ontology and the Cloud QoS Ontology.

1. **Cloud Service Ontology:** A *CloudService* (maps to *cloud_service_types* in the relational model in Figure 2) can be of one of the three types, Infrastructure-as-a-Service (IaaS), Platform-as-a-Service (PaaS) or Software-as-a-Service (SaaS). For the CloudRecommender system the cloud infrastructure layer (IaaS), providing concepts and relations that are fundamental to the other higher-level layers, is the one currently relevant. Cloud services in the IaaS layer can be categorised into: Compute, Network, and Storage services (see Table I).

2. **Cloud QoS ontology:** At the core of the Cloud QoS ontology is a taxonomy of *ConfigurationParameters* and *Metrics (Values)*, i.e. two trees formed using the RDF(s) subClassOf relation where an *Configuration Parameters*, for example, PriceStorage, PriceCompute, PriceDataTransferIn (Out) etc. and a Metric, for example, ProbabilityOfFailureOnDemand, TransactionalThroughput, are used in combination to define Cloud QoS capabilities (e.g. features, performance, costs, etc.). The resulting ontology is a (complex) directed graph where, for example, the Property *hasMetric* (and its inverse *isMetricOf*) is the basic link between the *ConfigurationParameters* and Metric trees. For the metrics part of the QoS, we reference existing QoS ontologies [17] whereas for the *ConfigurationParameters* concepts the ontology defines its independent taxonomy, but refers to external ontologies for existing definitions. Each configuration parameter (see Table I) has a name, and a value (qualitative or quantitative). The type of configuration determines the nature of service by means of setting a minimum, maximum, or capacity limit, or meeting certain value. For example, "RAM capacity" configuration parameter of a compute service can be set to the value 2GB

For our CloudRecommender service we implemented the Cloud Service Ontology in a relational model and the Cloud QoS ontology as configuration information as structured data (entities) (as shown in Figure 2), which can be queried using a SQL-based declarative language. We collected service configuration information from a number of public cloud providers (e.g., Windows Azure, Amazon, GoGrid, RackSpace, Nirvanix, Ninefold, SoftLayer, AT and T Synaptic, Cloud Central, etc.) to demonstrate the generic nature of the domain model with respect to capturing heterogeneous configuration (see Table II) information of infrastructure services. Our model is generic enough to capture all the existing cloud-based infrastructure services. The proposed model is flexible and extensible enough to accommodate new services with minimal changes to our implementation. In future work, we also intend to extend the model with capability to store PaaS and SaaS configurations.

| Service | Configurations Parameters | Range/possible values |
|---|---|---|
| | cores | >=1 |
| | speed | >0 |
| | RAM capacity | >0 |
| | local storage capacity | >=0 |
| | physical location of cloud | North America,South America,Africa,Europe,Asia,Australia |
| | cost per hour of usage per month | >=0 |
| | per period cost | >= 0 |
| | period length in days | > 0 |
| | overage cost | >= 0 |
| Compute | plan type | Pay As You Go, Prepaid |
| | network storage size min and max range | >=0 |
| | cost per GB per Month | >=0 |
| | location of host cloud | North America,South America,Africa,Europe,Asia,Australia |
| | type of requests | put,copy,post,list,get,delete,search |
| | cost of request | >=0 |
| | plan type | Pay As You Go, Reduced Redundancy |
| Storage | Reduced Redundancy storage cost | >=0 |
| Network | data transfer in cost, data transfer out cost | >=0 |

**Table I. Infrastructure service types and their configurations.**

Relationships between concepts representing services are carefully considered and normalized to avoid update anomalies. Services from various providers often have very different configurations and pricing models. Distinct and ambiguous terminologies are often used to describe similar configurations.

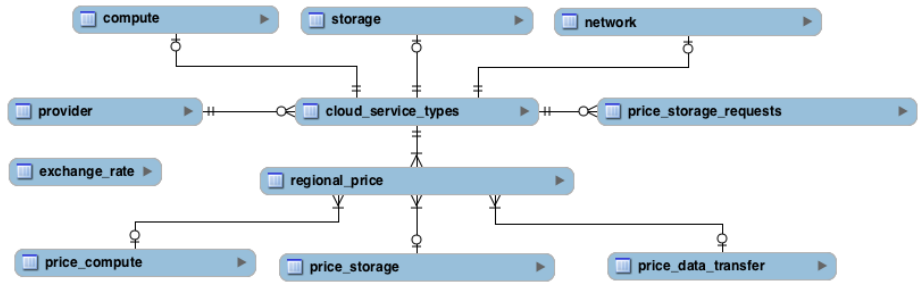

**Figure 2: Conceptual data model representing infrastructure service entities and their relationships.**

Regardless of how providers name their services, we categorize infrastructure services based on their basic functionality. Unit conversions were performed during instantiation of concepts. For example, an Amazon EC2 Micro Instance has 613 MB of memory which is converted to approximately 0.599 GB. Another example is the CPU clock speed. Amazon refers to it as "ECUs". From their documentation [10]: "*One EC2 Compute Unit provides the equivalent COMPUTE capacity of a 1.0-1.2 GHz 2007 Opteron or 2007 Xeon processor. This is also the equivalent to an early-2006 1.7 GHz Xeon processor referenced in our original documentation*". In 2007, AMD and Intel released both dual-core and quad-core models of the Opteron and Xeon chips, respectively. So it is obviously not clear what an Amazon EC2 Compute Unit compares to. To eliminate this ambiguity, we obtained the compute service clock speed by trying out the actual instance under Linux OS and run "more /proc/cpuinfo" on it. Table II depicts the configuration ambiguities of compute and storage services of different providers.

| Provider | Compute Terminology | Pay As You Go Unit | Number of Other Plans* | Storage Terminology | Pay As You Go Unit | Number of Other Plans* | Trail Period or Value |
|---|---|---|---|---|---|---|---|
| Windows Azure | Virtual Server | /hr | 1 | Azure Storage | /GB month | 1 | 90 day |
| Amazon | EC2 Instance | /hr | 2 | S3 | /GB month | 2 | 1 year |
| GoGrid | Cloud Servers | /RAM hr | 1 | Cloud Storage | /GB month | | |
| RackSpace | Cloud Servers | /RAM hr | | Cloud Files | /GB month | | |
| Nirvanix | | | | CSN | /GB month | | |
| Ninefold | Virtual Server | /hr | | Cloud Storage | /GB month | 1 | 50 AUD |
| SoftLayer | Cloud Servers | /hr | 1 | Object Storage | /GB | | |
| AT and T Synaptic | Compute as a Service | vCPU per hour + /RAM hr | | Storage as a Service | /GB month | | |
| Cloudcentral | Cloud Servers | /hr | | | | | |
| * Monthly/Quarterly/Yearly Plan, Reserve and Bidding Price Option | | | | | | | |

**Table II. Depiction of configuration heterogeneities in compute and storage services across providers. (Red) Blank cells in the table mean that a configuration parameter is not supported. Some providers offer their services under a different pricing scheme than pay-as-you-go. In Table II we refer to these schemes as other plans (e.g. Amazon Reduced redundancy, reserved price plans, GoGrid Pre-Paid plans).**

Another example of disparity between different cloud providers is the way in which "on Demand instances" are priced. GoGrid's plan, for example, although having a similar concept to Amazon's On Demand and Reserved Instance, gives very little importance to what type or how many of compute services a user is deploying. GoGrid charges users based on what they call RAM hours – 1 GB RAM compute service deployed for 1 hour consumes 1 RAM Hour. A 2 GB RAM compute service deployed for 1 hour consumes 2 RAM Hour. It is worthwhile mentioning that only Azure clearly states that one month is considered to have 31 days. This is important as the key advantage of the fine grained pay-as-you-go price model which, for example, should charge a user the same when they use 2GB for half a month or 1 GB for a whole month. Other vendors merely give a GB-month price without clarifying how short term usage is handled. It is neither reflected in their usage calculator. We chose 31 days as default value in calculation.

| Provider | Storage | Requests | | |
|---|---|---|---|---|
| | | Upload | Download | Other |
| Windows Azure | Azure Storage | storage transactions | storage transactions | |
| Amazon | S3 | PUT, COPY, POST, or LIST Requests | GET and all other Requests | Delete |
| GoGrid | Cloud Storage | Transfer protocols such as SCP, SAMBA/CIFS, and RSYNC | | |
| RackSpace | Cloud Files | PUT, POST, LIST Requests | HEAD, GET, DELETE Requests | |
| Nirvanix | CSN | | Search | |
| Ninefold | Cloud Storage | GET, PUT, POST, COPY, LIST and all other transactions | | |
| SoftLayer | Object Storage | Not Specified/Unknow | | |
| AT and T Synaptic | Storage as a Service | Not Specified/Unknow | | |

**Table III. Depiction of configuration heterogeneities in request types across storage services.**

Regarding storage services, providers charge for every operation that an application program or user undertakes. These operations are effected on storage services via RESTful or SOAP API. Cloud providers refer to the same set of operations with different names, for example Azure refers to storage service operations as transactions. Nevertheless, the operations are categorized into upload and download categories as shown in Table III. Red means an access fee is charged, green means the service is free, and yellow means it is not specified and usually can be treated as green/free of charge. To facilitate our calculation of similar and equivalent requests across multiple providers, we analyzed and pre-processed the price data, recorded it in our domain model and used a homogenized value in the repository (configuration management layer). For example, Windows Azure Storage charges a flat price per transaction. It is considered as transaction whenever there is a "touch" operation (a Create, Read, Update, Delete (CRUD) operation over the RESTful service interface) on any component (Blobs, Tables or Queues) of Windows Azure Storage.

For providers that offer different regional prices, we store the location information in the price table. If multiple regions have the same price, we choose to combine them. In our current implementation, any changes to existing configurations (such as updating memory size, storage provision etc.) of services can be done by executing customized update SQL queries. We also use customized crawlers to update provider information's periodically. However, in future work we will provide a RESTful interface and widget which can be used for automatic configuration updates.

## 2.2 Application Logic Layer

The request for service selection in CloudRecommender is expressed as SQL queries. The selection process supports an application logic that builds upon the following declarative constructs: criterion, views and stored procedures. The CloudRecommender builds upon SQL queries which are executed on top of the relational data model.

**Criterion**: Criterion is a quantitative or qualitative bound (minimum, maximum, equal) on the configuration parameters provided by a service. Cloud services' configuration parameters and their range/values listed in Table I form the basis for expressing selection goal and criteria (e.g., select a cheapest (goal) compute service where (criterion) 0<Ram<=20, 0<=local storage<=2040, number of hours to be used per month = 244). An example query is shown below in Fig 3:

```
27  SELECT *
28  FROM `compute_service_price`
29  left join compute_choice_criterias
30  on `Memory(GB)` >= compute_choice_criterias.ram_low
31  and `Memory(GB)` <= compute_choice_criterias.ram_high
32  and `Local Storage(GB)` >= compute_choice_criterias.local_storage_low
33  and `Local Storage(GB)` <= compute_choice_criterias.local_storage_high
34  where if(all_provider_considered, 1, find_in_set(`Provider Name`,provider_LIST))
```

**Figure 3: Example query in procedure.**

**Procedures:** We have implemented a number of customized procedures that automate the service selection process. A number of routines are prepared to process a user service selection request. List of inputs are stored in a temporary table to be passed into the procedures. As such, there is no limit to the size of the input list. Final results are also stored in temporary tables, which are automatically cleared after the expiration of user session.

| Notations | Meaning |
|---|---|
| $P = \{p_1, …, p_p\}$ | Set of p service providers |
| $R_{p_i} = \{r_{p_i,1}, …, r_{p_i,n}\}$ | Regions of provider $p_i$ |
| $CS = \{cs_1, …, cs_n\}$ | Set of n compute services |
| $SS = \{ss_1, …, ss_m\}$ | Set of m storage services |
| $TS = \{ts_1, …, ts_o\}$ | Set of o network (data transfer) services |
| $t_{s_i,j}$ | j-th price tier for a cloud service $s_i \in CS \cup SS \cup TS$ |
| $CR_{s_i} = \{cr_{s_i,1}, …, cr_{s_i,n}\}$ | Set of criteria related to service $s_i \in CS \cup SS \cup TS$ |
| Query | A service selection query |
| N | Number of rows in a relational entity |
| M | Number of column in a relational entity |

**Table IV: CloudRecommender Model Parameters.**

**Controller:** The controller supports enforcement of criteria and dynamically generates SQL queries which fulfill a given selection preference stated by the user. Due to space considerations we are not able to depict the complete algorithm, but Figure 4 shows the selection logic in a simplified diagram. Next we explain the basic steps which are executed for resolving a service selection request.

1. Basic validation is preformed on user inputs at the controller, appropriate errors are returned accordingly.
2. Depending on user's requirements, process 3.2 or 3.3 may not happen. This is why they are shown as dotted lines, i.e. user can query storage or compute only IaaS services. But data transfer parameters have to be set as user will definitely transfer data in and out of the compute or storage service.
3. Multiple temporary tables are created during the process so intermediate results (i.e. selection details of the final recommendation) can be fetched later as needed.
4. It is possible for a user to choose multiple compute services each with different criteria. (E.g. they may have 10 set of requirements, and choose 10 instances for each.) So in process 5, query with different number of join operations are dynamically constructed.
5. Currency conversions are performed before costs are compared.

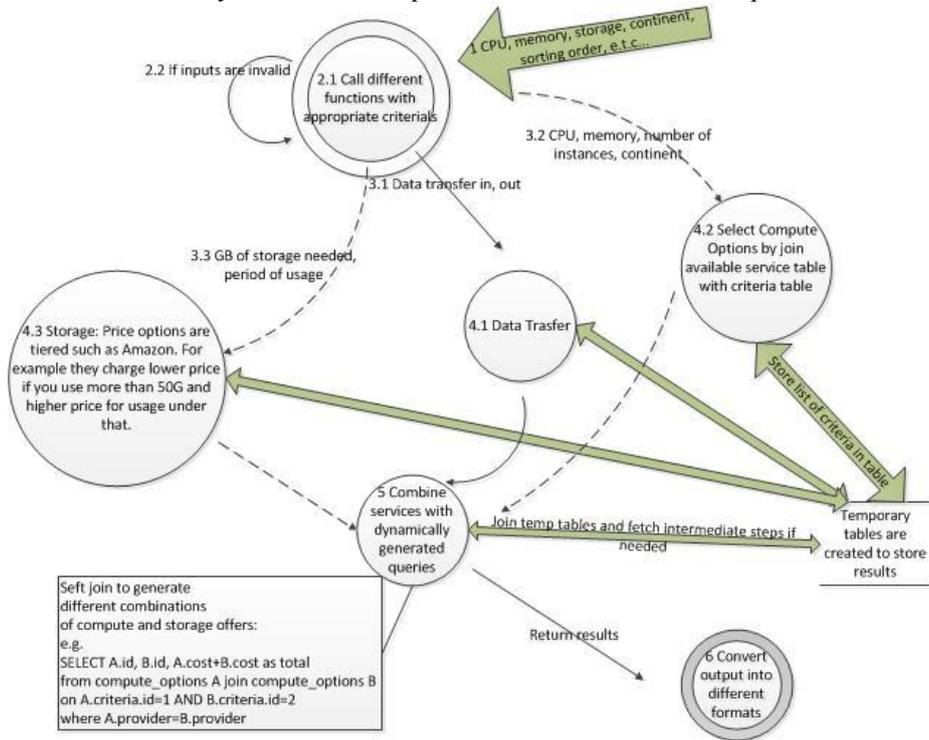

**Figure 4: Service Selection logic.**

**Computational Complexity of Service Selection Logic**

We will discuss the computational complexity of our service selection logic next. For p providers each with $cs_i$ (compute) + $ss_i$ (storage) + $ts_i$ (network) services, the selection logic has to consisder $\sum_{i=1}^{p} cs_i \times ss_i \times ts_i$ choices. We give the detailed discussion of model parameters in Table IV. We can nomally reduce the number of options significantly in the early stage if a user has strict requirements. In the worst case scenario, the logic needs to compute a full cross join (cartesian product). The number of choices varies depending on the number of regions ($R_{p_i}$) with different prices offered by each provider ($r_i$), and the number of different price tier ($t_i$) for each service (Price tier example: AWS S3 charges $0.125 per GB for the first 1 TB / month of usage, $0.093 for the next 49 TB, etc.). Depending on the estimated usage, the larger the usage, the more price tiers will be involved. Let us assume that each provider offers approximately the same service in each region to simplify the derivation of the computational complexity. As such, the total number of offers can be represented in a more detailed formula:

$$\sum_{i=1}^{p} \left(\sum_{l=1}^{cs_i} t_l\right) \times \left(\sum_{m=1}^{ss_i} t_m\right) \times \left(\sum_{n=1}^{ts_i} t_n\right) \times r_i$$

The queries of the selection logic works as follows. After filtering out criteria-violating services, resulting services are combined via JOIN operation(s) with final costs calculated. In worst case scenario where a few or no criteria are defined, the combination of the services is a full CROSS JOIN over all existing services. Therefore, the selection queries, to our best knowledge, have the upper bound computational complexity of

$$O_{query}(|cr_{compute}| \sum_{i=1}^{p} |cs_i| \times |cr_{storage}| \sum_{i=1}^{p} |ss_i| \times |cr_{network}| \sum_{i=1}^{p} |ts_i|)$$

where $cr$ are criteria and $cs, ss$ and $ts$ are pre-computed views with a singular effort to create the views from JOIN statements. However, in case the database system lacks support for cached views in a worst case the effort multiplies with the effort of the views' JOIN. Modern database can use HASH JOIN O(N + M) and MERGE JOIN O(N*Log(N) + M*Log(M)) which are faster than O(N * M).

### 2.3 Widget Layer

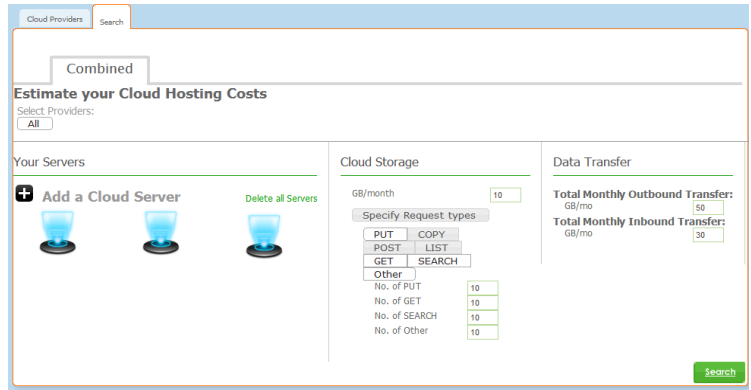

**Figure 5: Screen shot of the widget.**

This layer features rich set of user-interfaces (see Fig 5 and Fig 9) that further simplify the selection of configuration parameters related to cloud services. This layer encapsulates the user interface components in the form of four principle widgets including: Compute, Storage, Network, and Recommendation. The selection of basic configuration parameters related to compute services including their RAM capacity, cores, and location can be facilitated through the Compute widget. It also allows users to search compute services by using regular expressions, sort by a specific column etc. Using the Compute widget, users can choose which columns to display and rearrange their order as well. The Storage widget allows users to define configuration parameters such as storage size and request types (e.g., get, put, post, copy, etc.). Service configuration parameters, such as the size of incoming data transfer and outgoing data transfer can be issued via the Network widget. Users have the option to select single service types as well as bundled (combined search) services driven by use cases. The selection results are displayed and can be browsed via the Recommendation widget (not shown in Fig 5).

## 3  Experiments and Evaluation

In this section, we present the experiments and evaluation that we undertook.

**Experiment Setup:** We hosted our CloudRecommender system instance on Amazon EC2 using a standard small instance in the US-west location. By default, the small instance has the following hardware configuration: 1.7 GB of main memory, 1 EC2 Compute Unit, 160 GB of local instance storage, and a 32-bit platform with an Ubuntu 10.04 operating system. We populated CloudRecommender with infrastructure service configuration information related to Amazon, Azure, GoGrid, RackSpace, Nirvanix, Ninefold, SoftLayer, AT & T Synaptic, and Cloud Central (an Australian provider).

**CloudRecommender: service selection test case**

Figure 6: Service selection criteria set by business analyst.

In our infrastructure service selection scenario, we revisit the example of a medium scale enterprise we explained earlier. The enterprise wants to migrate its data to the cloud with the aim of storing and sharing it with other branches through public cloud storage (note that security issues are dealt within the enterprise applications). At this stage, we assume the business analyst of the enterprise has a good estimation of the data storage and transfer (network in/network out) requirements. By using CloudRecommender, the analyst would like to find out which of the public cloud providers would be most cost-effective in regards to data storage and transfer costs. For this selection scenario, the analyst inputs the following anticipated usage information for the storage and network services: (i) Data size of 50 GB, 1000 copy requests and 5000 get requests and (ii) data transfer in size of 10 GB and data transfer out size of 50 GB.

As shown in Fig 6, the analyst specifies service selection criteria via the storage and network widgets. Programmatically, the above request can also be submitted via the RESTful service interface of the CloudRecommender as shown below in Fig 7.

```
Call
localhost/cloud_demo_1_1/api/cost/storage?
media_type=xml¤cy=AUD&storage=50&duration=31&data_upload_size=50&data_download_size=10©=1000&get=5000
```

**Figure 7: An Example REST call.**

Once this selection request is submitted, the controller validates and parameterizes the criteria (user inputs). Though not shown in Fig 6, the business analyst also has the option to express whether the selection criteria should be evaluated against all the available cloud providers or only the selected ones (e.g., Amazon, Azure, and GoGrid only). As mentioned earlier, the application logic layer implements specialized views and procedures for evaluating different service selection scenarios. CloudRecommender captures and inserts multiple storage and network service selection criteria into specialized views called "storage_selection_criteria" and "network_selection_criteria". Aforementioned views are then joined against the "storage_service_price" and "network_service_price" views for estimating the cost of using the combined cloud services.

| Provider Name | region_name | storage_cost | cost_data_in | cost_data_out | storage_dataTransfer_cost |
|---|---|---|---|---|---|
| SoftLayer | Any | 6.00000000 | 0.000 | 1.000 | 7 |
| Windows Azure | North America and Europe | 7.00000000 | 0.000 | 1.200 | 8.206 |
| Amazon | Asia Pacific(Tokyo) | 6.76500000 | 0.000 | 0.000 | 8.589 |
| Windows Azure | Asia Pacific Region | 7.00000000 | 0.000 | 1.900 | 8.906 |
| Ninefold | Any | 4.60000000 | 0.000 | 9.000 | 13.606 |
| AT and T Synaptic | Any | 10.00000000 | 5.000 | 1.000 | 16 |
| AT and T Synaptic | Any | 10.00000000 | 5.000 | 1.000 | 16 |
| AT and T Synaptic | Any | 10.00000000 | 5.000 | 1.000 | 16 |
| AT and T Synaptic | Any | 12.50000000 | 5.000 | 1.000 | 18.5 |
| Nirvanix | Any | 12.50000000 | 5.000 | 1.500 | 19 |
| Nirvanix | Any | 12.50000000 | 10.000 | 1.500 | 24 |
| Nirvanix | Any | 12.50000000 | 15.000 | 1.500 | 29 |

**Table V. Storage and network services recommendations for the business analyst selection use case.**

Table V shows the result of the RESTfull call or the selection scenario depicted in Fig 6. Results are sorted into increasing total cost order (i.e. "storage_dataTransfer_cost" column). "Any" means the provider offers the same price for all of its regions. In the case of SoftLayer, it charges the same price in all regions. In the case of Amazon AWS, since it offers different prices for different regions, the enterprise may be able to consume the same service with a cheaper price in a different region. The base currency is USD, but since in the call the analyst had specified "currency=AUD", the result shown below is in AUD accurate to one decimal place. Fig 8 shows another example which selects compute, storage, and network using the RESTful API. The selection criteria include 1 compute service instance (shown as "n" in Fig 8): 0<Ram<=69, 0<=local storage<=2040, number of hours to be used per month 744. The selection results are displayed at the end of Fig 8.

Due to high inter-cloud data transfer cost overhead and communication delay, our recommender logic does not consider the combination of services from multiple providers. For example, the CloudRecommender will not select and combine compute service from Amazon with storage service from Azure. In future work we intend to integrate a run-time network QoS statistics (e.g., inter-cloud latency, inter-cloud upload speed, and inter-cloud download speed) information to the CloudRecommender. Similarly, some provider charge for data transfer across their own services that are hosted at different regions. For example, data transferred between Amazon services in different regions are charged as "Internet Data Transfer" on both sides of the transfer. We currently choose to put all services in the same region. In future we will extend our recommendation logic to allow users to choose between different regions for each service type (if necessary). Additionally providers often offer discounted price for higher usage, keeping all data together means higher usage which can consume a cheaper price tier. Network services are always bundled with either compute or storage service as it is impossible to consume other services without incurring network costs.

Figure 8. Result of selection process for Compute, Storage and network services

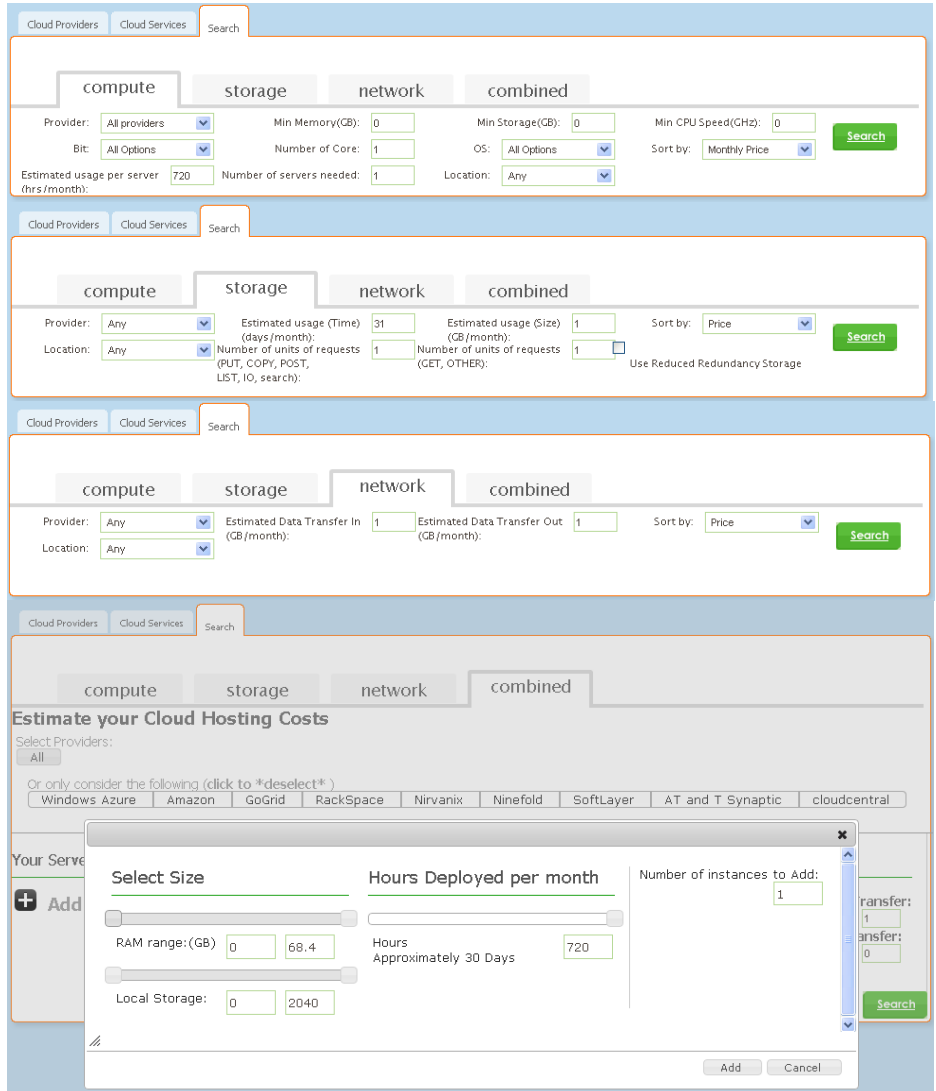

Figure 9: Screen shots of Compute, Storage, Network and the combined service selection widgets.

## 4 Related Work

Prior to CloudRecommender, there have been a variety of systems that use declarative logic-based techniques for managing resources in distributed computing systems. The focus of the authors in work [4] is to provide a distributed platform that enables cloud providers to automate the process of service orchestration via the use of declarative policy languages. The authors in [5] present an SQL-based decision query language for providing a high-level abstraction for expressing decision guidance problems in an intuitive manner so that database programmers can use mathematical programming technique without prior experience. We draw a lot of inspiration from the work in [6] which proposes a data-centric (declarative) framework to orchestrate infrastructure services. The goal of this work is to improve SLA fulfilment ability of cloud service providers. COOLDAID [7] presents a declarative approach to manage configuration of network devices and adopts a relational data model and Datalog-style query language. NetDB [8] uses a relational database to manage the configurations of network devices. However, NetDB is a data warehouse, not designed for cloud service selection and composition. Puppet [9] manages the configuration of data-centre resources using a custom and user-friendly declarative language for service configuration specifications. Puppet simplifies the management of data centre resources for providers. Though branded calculators are available from individual cloud providers, such as Amazon [14], Azure [15], and GoGrid, for calculating service leasing cost, it is not easy for users to generalize their requirements to fit different service offers (with various quota and limitations) let alone computing and comparing costs. Some of the recent research such as [11] has focused on cloud storage service (IaaS level) representation based on an XML schema. However, the proposed declarative model is preferable over hard coding the sorting and selection algorithm (as used in [11]) as it allows us to take the advantage of optimized SQL operations (e.g. select and join).

In contrast to the aforementioned systems, CloudRecommender is designed with a different application domain – one that aims to apply declarative (SQL) and widget programming technique for solving the cloud service configuration selection problem. Facing a new challenge of handling heterogeneous service configuration and naming conventions in cloud computing, CloudRecommender also defines and uses a unified domain model.

## 5 Conclusion and Future Work

In this paper, we proposed a declarative system (CloudRecommender) that transforms the cloud service configuration selection from an ad-hoc process that involves manually reading the provider documentations to a process that is structured, and to a large extend automated. Although we believe that CloudRecommender leaves scope for a range of enhancements, yet provides a practical approach. We have implemented a prototype of CloudRecommender and evaluated it using an example selection scenario. The prototype demonstrates the feasibility of the CloudRecommender design and its practical aspects.

Our future work includes: (1) extending the CloudRecommender to support the selection of more cloud service types such as PaaS services (e.g., database server, web server, etc.) to further validate our hypothesis and explore new opportunities; (ii) exploring integration of cloud service benchmarking databases such as CloudHarmony to CloudRecommender for facilitating run-time selection based on dynamic QoS information including throughput, latency, and utilization; and (iii) deploying and evaluating the CloudRecommender as a REST service so that it can be easily integrated to any existing cloud service orchestration systems.

**Acknowledgments.** Initial research on the infrastructure service data models was done when Dr. Rajiv Ranjan was employed at University of New South Wales (UNSW) on a strategic eResearch grant scheme.